\documentclass[lettersize,journal]{IEEEtran}

\usepackage{amsmath,amssymb,amsfonts,amsthm,mathtools,bm}
\usepackage{cite} 
\usepackage{CJKutf8}
\usepackage{algorithm}
\usepackage{algorithmic}
\usepackage{graphicx}
\usepackage{stfloats}
\usepackage{booktabs}
\usepackage{caption}
\usepackage{titlesec}
\usepackage{indentfirst}

\usepackage{epsfig,subfigure,upgreek}
\usepackage{hyperref}
\usepackage{xcolor}  
\usepackage{etoolbox}

\titlespacing*{\section}{0pt}{0.8ex plus 0.2ex minus 0.2ex}{0.6ex plus 0.1ex}
\titlespacing*{\subsection}{0pt}{0.6ex plus 0.2ex minus 0.2ex}{0.4ex plus 0.1ex}
\hypersetup{
	colorlinks=true,
	linkcolor=black,         
	citecolor=black,         
	urlcolor=black          
}

\ifCLASSINFOpdf

\else

\fi

\begin{document}
	
\title{A Low-Complexity Message-Passing Receiver for FTN-WAFT under Nonlinear Distortion}

\author{
% \IEEEauthorblockN{Xianle Dai}
\IEEEauthorblockN{Xianle Dai, Qu Luo,~\IEEEmembership{Member,~IEEE,} Jianguo Li,~\IEEEmembership{Member,~IEEE,} Bowei Xue, Kai Yang,~\IEEEmembership{Member,~IEEE,}  Luciano Leonel Mendes,~\IEEEmembership{Member,~IEEE,} and Pei Xiao,~\IEEEmembership{Senior Member,~IEEE}}
\thanks{This work was supported by the National Natural Science Foundation of China under Grant 62401044. (\textit{Corresponding author: Jianguo Li, Kai Yang.})}
\thanks{Xianle Dai and Kai Yang are with the School of Information and Electronics, Beijing Institute of Technology, Beijing 100081, China (e-mail: \{daixl212, yangkai\}@bit.edu.cn).}% <-this % stops a space
\thanks{Qu Luo and Pei Xiao are with the 5G/6G Innovation Centre, University of Surrey, GU2 7XH Guildford, U.K. (e-mail: \{q.u.luo, p.xiao\}@surrey.ac.uk).}
\thanks{Jianguo Li and Bowei Xue are with the School of Cyberspace Science and Technology, Beijing Institute of Technology, Beijing 100081, China (e-mail: \{jianguoli, boweixue\}@bit.edu.cn).}
\thanks{Luciano Leonel Mendes is with National Institute of Telecommunications, Santa Rita do Sapucai, 37536-001, Brazil (e-mail: lucianol@inatel.br).}
\vspace{-0.5cm}
}

\maketitle
\begin{abstract}
This letter proposes a faster-than-Nyquist (FTN) weighted affine Fourier transform (WAFT) transmission scheme, termed FTN-WAFT, which combines the peak-to-average power ratio (PAPR) flexibility of WAFT with FTN time compression to improve spectral efficiency and power amplifier (PA) efficiency over high-mobility channels. However, operating the PA at a reduced input back-off (IBO) inevitably introduces nonlinear distortion, which further couples with FTN-induced inter-symbol interference (FTN-ISI) and doubly selective channel dispersion. To address this practical impairment, a multilayer message-passing (MLMP) receiver aided by PA interference cancellation (PA-IC), termed PA-IC-MLMP, is proposed. The proposed receiver reconstructs the dominant PA distortion from soft symbol estimates, cancels the channel-propagated nonlinear interference, and performs MLMP over the WAFT, FTN, channel, and constellation layers. Simulation results show that the proposed receiver improves the bit error rate (BER) and effective throughput under nonlinear PA constraints while maintaining low complexity.
\end{abstract}


% ============================================================================
\section{Introduction}
% ============================================================================

Emerging applications in the upcoming sixth generation (6G) wireless networks, such as connected autonomous vehicles, drone swarms, and low-Earth-orbit satellites, impose a  strong demand  for  reliable communication over high-mobility channels \cite{Zhang2019SixG}. The legacy orthogonal frequency-division multiplexing (OFDM) may be ineffective in such scenarios due to inter-carrier interference and its high peak-to-average power ratio (PAPR) \cite{luo2026towards}.  
To enhance transmission robustness over high-mobility channels, emerging waveforms have been investigated, such as affine frequency division multiplexing (AFDM) \cite{Bemani2023AFDM}, orthogonal chirp division multiplexing (OCDM) \cite{Ouyang2016OCDM}, and orthogonal time-frequency space (OTFS) modulation \cite{Raviteja2018OTFSMP}.
More recently, the weighted affine Fourier transform (WAFT) was introduced to combine single-carrier (SC) and chirp multi-carrier (CMC) components within a unified transform framework \cite{Li2024WAFT}. 
By tuning the WAFT order, the relative weights of the two components can be adjusted, providing a flexible waveform structure for PAPR control and high-mobility channel adaptation.

A complementary approach to improving spectral efficiency is faster-than-Nyquist (FTN) signaling, where symbols are transmitted faster than the Nyquist rate without increasing the occupied bandwidth \cite{DaiXianle_AFDMFTN,Bedeer2017QAMFTN,Yang2026BitMappingFTN,Yang2025Improved5GLDPCFTN,Yang2026GOrientedFTN}. Specifically, related studies have considered high-order quadrature amplitude modulation (QAM) \cite{Bedeer2017QAMFTN} and LDPC-coded FTN signaling \cite{Yang2026GOrientedFTN}. By enhancing pulse overlap, FTN may reshape the waveform envelope to reduce PAPR. \cite{Peng2018MFTNPAPR}. Combining FTN with WAFT therefore exploits both spectral-efficiency and transform-domain degrees of freedom, while FTN-induced inter-symbol interference (FTN-ISI) makes receiver design more challenging.

In practical transceivers, AFDM and related chirp-domain waveforms are also affected by hardware constraints. 
Existing studies have considered low-bit analog-to-digital converters \cite{mengPerformanceAnalysisAFDM2026}, practical pulse shaping and hardware impairments \cite{mirabellaContinuousTimeAnalysisAFDM2026}, and PAPR control \cite{liuAFDMTransceiverOptimization2026}, which directly affect quantization robustness, waveform envelope, and PA operating efficiency. 
A high-PAPR signal requires a larger input back-off (IBO) for near-linear PA operation, which reduces PA efficiency. 
With limited back-off, PA-induced distortion interacts with FTN-ISI and doubly selective channel dispersion, leading to a nonlinear detection problem \cite{Ronnow2019Bussgang}. 
Since OFDM-oriented distortion-compensation methods do not directly exploit the layered WAFT, FTN, and channel structure \cite{Sun2021OFDMRecovery}, waveform design and nonlinear receiver processing should be jointly considered for FTN-WAFT under practical hardware constraints.
 
Motivated by these observations, this letter first proposes an FTN-WAFT waveform for spectrally efficient communication over high-mobility channels. %To address the PA nonlinearity in practical transmitters, a PA-IC-MLMP receiver is then developed to mitigate the nonlinear distortion at the receiver side. 
The main contributions are summarized as follows.   First, an FTN-WAFT waveform is designed by combining WAFT-based hybrid carrier modulation with FTN signaling, where the transform order, chirp rate, and compression factor provide multiple degrees of freedom for PAPR reduction, spectral-efficiency improvement, and reliable detection in high-mobility channels.
Second, a PA-IC-MLMP receiver is developed to jointly address PA distortion, FTN-ISI, and channel dispersion under reduced IBO. The receiver reconstructs the dominant PA distortion from soft symbol estimates, cancels the channel-propagated nonlinear interference, and performs MLMP over the WAFT, FTN, channel, and constellation layers.
Simulation results show that FTN-WAFT reduces PAPR and supports adaptive IBO, while the proposed receiver converts this waveform advantage into BER and effective throughput gains.

The letter is organized as follows. Section~\ref{sec:sysmodel} presents the FTN-WAFT system model under nonlinear distortion. Section~\ref{sec:receiver} develops the proposed PA-IC-MLMP receiver and analyzes its complexity. Section~\ref{sec:sims} provides simulation results, and Section~\ref{sec:conclusion} concludes this letter.

% ============================================================================
\section{System Model}
\label{sec:sysmodel}
\begin{figure*}[!t] \centerline{\includegraphics[width=0.9\textwidth]{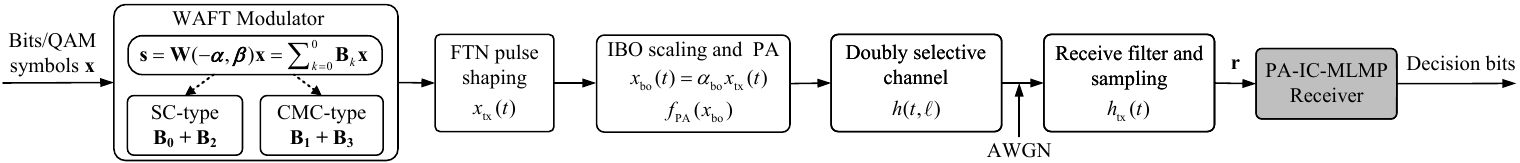}}
	\captionsetup{justification=raggedright,singlelinecheck=false}
	\caption{Transceiver block diagram of the proposed FTN-WAFT system with nonlinear distortion.}
	\label{fig1}
	\vspace*{-0.4cm}
\end{figure*}

% ============================================================================
This section describes the system model of the proposed FTN-WAFT under nonlinear distortion, as shown in Fig.~\ref{fig1}.

\subsection{WAFT Framework}
We first introduce the WAFT kernel \cite{Li2024WAFT}. Define the $N\times N$ chirp-phase matrix $\mathbf \Lambda_\beta \triangleq \mathrm{diag}\{e^{j2\pi\beta n^2}\}_{n=0}^{N-1}$, where $\beta$ denotes the chirp rate. With weighting order $\alpha$, the WAFT matrix is given by \cite{Li2024WAFT}
 \begin{equation}
	\scalebox{0.87}{$%
		\begin{aligned}
			\mathbf W(\alpha, \beta) = \sum_{i=0}^{3} \omega_i(\alpha)\, \mathbf {\Lambda}_\beta^H \mathbf F^i \mathbf {\Lambda}_\beta,
		\end{aligned}
	$}
	\label{eq:011}   
\end{equation}
where $\mathbf F$ is the normalized discrete Fourier transform (DFT) matrix with $(m,n)$th entry $e^{-j2\pi mn/N}/\sqrt{N}$, and
 \begin{equation}
	\scalebox{0.85}{$%
		\begin{aligned}
			\omega_i(\alpha) = \cos\left(\frac{(\alpha-i)\pi}{4}\right)
			\cos\left(\frac{2(\alpha-i)\pi}{4}\right)
			\exp \left(j\frac{3(\alpha-i)\pi}{4}\right),
		\end{aligned}
		$}
	\label{eq:02}   
\end{equation}
and the coefficients satisfy $\sum_{i=0}^{3}|\omega_i(\alpha)|^{2}=1$. 
The unitary matrix $\mathbf{W}(\alpha,\beta)$ satisfies $\mathbf{W}(0,\beta)=\mathbf{I}_N$ and $\mathbf{W}(1,\beta)=\mathbf A$, where  $\mathbf{A}\triangleq \mathbf{\Lambda}_\beta^{H}\mathbf{F}\mathbf{\Lambda}_\beta$.
In particular, compared with AFDM, WAFT uses the same chirp rate for both matrices.

Let $\mathbf x=[x_0,\ldots,x_{N-1}]^T$ denote normalized QAM symbols.  
    The resulting time-domain WAFT-based hybrid carrier (WAFT-HC) vector is given by 
 \begin{equation}
	\scalebox{0.87}{$%
		\begin{aligned}
  			\mathbf{s} = \mathbf W(-\alpha, \beta) \mathbf{x}.
		\end{aligned}
		$}
	\label{eq:03}   
\end{equation}
A cyclic prefix (CP) is applied  before transmission, and its expression is omitted for simplicity.

\subsection{FTN-WAFT under Nonlinear Distortion}
FTN signaling compresses the symbol interval from $T_s$ to $\xi T_s$ with $\xi\in(0,1]$, thereby improving spectral efficiency at the cost of controlled FTN-ISI. After FTN pulse shaping is applied to the WAFT-domain signal $\mathbf s$ with symbol interval $\xi T_s$, the baseband waveform is generated as
\begin{equation}
	\scalebox{0.87}{$%
		\begin{aligned}
  			x_{\mathrm{tx}}(t) = \sum_{n=0}^{N-1}  s_n h_{\rm{tx}}(t - n\xi T_s),
		\end{aligned}
	$}
\label{eq:031}   
\end{equation}
where $s_n$ is the $n$th entry of $\mathbf s$, and $h_{\rm{tx}}(t)$ denotes the transmit pulse designed for the nominal Nyquist interval $T_s$.
The waveform then propagates through a doubly selective channel  modeled by
\begin{equation}
	\scalebox{0.87}{$%
		\begin{aligned}
			h(t,\ell) = \sum_{i=0}^{L-1}h_ie^{-j2\pi f_it}\delta(\ell-l_i),
		\end{aligned}
		$}
	\label{eq:04}
\end{equation}
where $L$ is the path number, and $h_i$, $f_i$, and $l_i$ denote the path coefficient, Doppler shift, and integer delay, respectively.

At the receiver, $h_{\rm rx}(t)$ denotes the receive filter designed with an orthogonality interval of $\xi T_s$ \cite{prljaReducedComplexityReceiversStrongly2012}. With the transmit and receive pulse correlation $\rho(t)=\int_{-\infty}^{\infty}h_{\rm tx}(\tau)h_{\rm rx}^{*}(\tau-t)\,{\rm d}\tau$, the elements of the FTN-ISI matrix are given by
\begin{equation}
	\scalebox{0.87}{$%
		\begin{aligned}
			\mathbf G[m,n] = \rho\bigl((m-n)\xi T_s\bigr), \, 0\le m,n \le N-1.
		\end{aligned}
		$}
	\label{eq:05}
\end{equation}
Let $A_{\rm sat}$ denote the PA saturation amplitude, and define the IBO as the ratio of the input saturation power to the average PA input power. For a prescribed IBO, the FTN-shaped waveform is scaled before PA amplification as
\begin{equation}
	\scalebox{0.87}{$%
		\begin{aligned}
			x_{\rm bo}(t)=\gamma_{\rm bo}x_{\mathrm{tx}}(t),
		\end{aligned}
		$}
	\label{eq:pa_input}
\end{equation}
where $\gamma_{\rm bo}=A_{\rm sat}/\sqrt{10^{{\rm IBO}/10}P_{\rm tx}}$ is the transmit scaling coefficient, and $P_{\rm tx}=\mathbb E\{|x_{\mathrm{tx}}(t)|^2\}$ denotes the average power of $x_{\mathrm{tx}}(t)$.
The resulting PA input then passes through the solid-state power amplifier (SSPA) model, i.e., 
\begin{equation}
	\scalebox{0.87}{$%
		\begin{aligned}
			f_{\mathrm{PA}}(x)
			=
			\frac{x}{\left(1 + (|x|/A_{\mathrm{sat}})^{2p}\right)^{1/(2p)}},
		\end{aligned}
		$}
	\label{eq:PA}
\end{equation}
where $p$ is the smoothness parameter. For receiver design, the PA output is separated into a useful linear component and an uncorrelated distortion term through Bussgang's theorem as
\begin{equation}
	\scalebox{0.87}{$%
		\begin{aligned}
			f_{\mathrm{PA}}(x_{\mathrm{bo}}(t))
			=
			\kappa_{\mathrm{bo}}x_{\mathrm{bo}}(t)+d(t)
			=
			\tilde{\kappa}x_{\mathrm{tx}}(t)+d(t),
		\end{aligned}
		$}
	\label{eq:110}
\end{equation}
where $\tilde{\kappa}=\kappa_{\mathrm{bo}}\gamma_{\mathrm{bo}}$ is the effective Bussgang gain, and $d(t)$ denotes the nonlinear distortion uncorrelated with $x_{\mathrm{tx}}(t)$. $\tilde{\kappa}$ and the PA parameters in \eqref{eq:PA} are generally assumed to be known at the receiver. For $P_{\mathrm{act}}=\mathbb E\{|x_{\rm bo}(t)|^2\}$, the Bussgang coefficient is given by
\begin{equation}
	\scalebox{0.87}{$%
		\begin{aligned}
			\kappa_{\mathrm{bo}}
			=
			\frac{1}{P_{\mathrm{act}}}
			\int_0^\infty
			f_{\mathrm{PA}}(r) r
			\frac{2r}{P_{\mathrm{act}}}
			e^{-r^2/P_{\mathrm{act}}}
			\mathrm{d}r.
		\end{aligned}
		$}
	\label{eq:11}
\end{equation}

Assume an FTN truncation length of $N_I$ is applied, i.e., 
$\mathbf{G}[m,n]=0$ for $|m-n|>N_I$.
Assuming that the CP length is larger than  the maximum channel delay, the
noiseless received signal after channel propagation can be expressed as
$\mathbf{H}\mathbf{G}\mathbf{s}$.
Combining this factorized model with the PA decomposition in \eqref{eq:110} yields
\begin{equation}
	\scalebox{0.87}{$%
		\begin{aligned}
			\mathbf r
			=
			\tilde{\kappa}\mathbf H\mathbf G\mathbf W(-\alpha,\beta)\mathbf x
			+
			\mathbf d_{\rm eq}
			+
			\mathbf n,
		\end{aligned}
		$}
	\label{eq:received_pa}
\end{equation}
where $\mathbf H$ is the equivalent discrete matrix representation of $h(t,\ell)$, 
$\mathbf d_{\rm eq}$ denotes the equivalent PA distortion after channel propagation and receive filtering whose $m$th entry is $[d_{\rm eq}]_m=\int\left(\sum_{i=0}^{L-1}h_i e^{-j2\pi f_i t}d(t-l_i)\right)h_{\rm rx}^*(t-m\xi T_s)\,{\rm d}t$ and $\mathbf n\sim\mathcal{CN}(\mathbf 0,\sigma^2\mathbf I_N)$ denotes the sampled AWGN after receive filtering that is orthogonal at intervals of $\xi T_s$.

\section{Proposed PA-IC-MLMP Receiver}
\label{sec:receiver}
\begin{figure}[t]
	\centerline{\includegraphics[width=0.9\columnwidth]{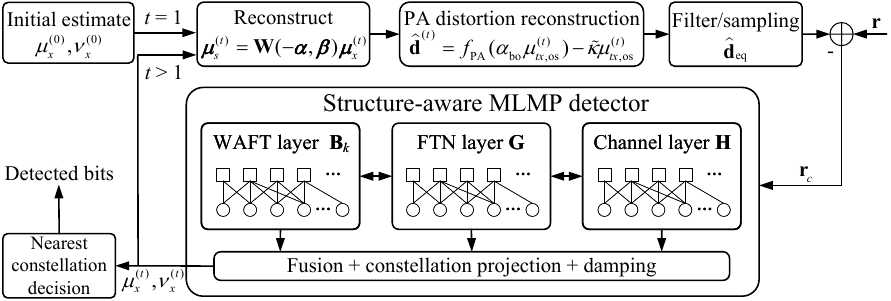}}
	\caption{Structure of the proposed PA-IC-MLMP receiver for FTN-WAFT with nonlinear distortion.}
	\label{fig2}
	% \vspace{-0.2cm}
\end{figure}

This section develops the PA-IC-MLMP receiver for FTN-WAFT under nonlinear distortion. The PA-IC-MLMP receiver combines PA-distortion reconstruction, layered message passing, and constellation projection, as illustrated in Fig.~\ref{fig2}.
\subsection{Iterative PA-IC-MLMP Detection}
In each iteration, the PA distortion is reconstructed from the current  waveform estimate and canceled after its propagation through the channel.

For compact notation, the inverse WAFT modulation matrix is decomposed as $\mathbf W(-\alpha,\beta) = \sum_{k\in\mathcal K}\mathbf B_k$,
where $\mathcal K=\{0,1,2,3\}$ and $\mathbf B_k=\omega_k(-\alpha)\mathbf\Lambda_\beta^H\mathbf F^k\mathbf\Lambda_\beta$.
The resulting observation is then processed by MLMP, where the WAFT, FTN, and channel operations are represented by $\{\mathbf B_k\}_{k\in\mathcal K}$, $\mathbf G$, and $\mathbf H$, respectively.
The linear part of the received signal can then be written as
\begin{equation}
	\scalebox{0.87}{$%
		\begin{aligned}
			\tilde{\kappa}\mathbf H\mathbf G\mathbf W(-\alpha,\beta)\mathbf x
			=
			\sum_{k\in\mathcal K}
			\tilde{\kappa}\mathbf H\mathbf G\mathbf B_k\mathbf x.
		\end{aligned}
		$}
	\label{eq:branch_signal}
\end{equation}
The iterative detection procedure is described as follows.

\textbf{Step 1: PA distortion cancellation.}
At the $t$th iteration, let $\boldsymbol{\mu}_x^{(t)}$ and $\boldsymbol{\nu}_x^{(t)}$ denote the propagated  mean and variance, respectively. The current symbol mean is first mapped through inverse WAFT modulation as $\boldsymbol{\mu}_s^{(t)}=\mathbf W(-\alpha,\beta)\boldsymbol{\mu}_x^{(t)}$.
The oversampled PA distortion is reconstructed as
\begin{equation}
	\scalebox{0.87}{$%
		\begin{aligned}
			\hat{\mathbf d}^{(t)}
			=
			f_{\rm PA}\left(\gamma_{\rm bo}\boldsymbol{\mu}_{\rm tx,os}^{(t)}\right)
			-
			\tilde{\kappa}\boldsymbol{\mu}_{\rm tx,os}^{(t)},
		\end{aligned}
		$}
	\label{eq:dhat}
\end{equation}
where $\boldsymbol{\mu}_{\rm tx,os}^{(t)}$ is obtained by applying FTN pulse shaping to $\boldsymbol{\mu}_s^{(t)}$ and oversampling the resulting soft waveform. After channel propagation, receive filtering, and sampling, the estimated distortion, denoted as $\hat{\mathbf d}_{\rm eq}^{(t)}$, is subtracted from the received signal as $\mathbf r_{\rm c}^{(t)} = \mathbf r - \hat{\mathbf d}_{\rm eq}^{(t)}$, while any residual nonlinear distortion due to parameter mismatch is absorbed into the residual disturbance variance.
The residual observation is then approximated as
\begin{equation}
	\scalebox{0.87}{$%
		\begin{aligned}
			\mathbf r_{\rm c}^{(t)}
			=
			\tilde{\kappa}\mathbf H\mathbf G\mathbf W(-\alpha,\beta)\mathbf x
			+
			\boldsymbol{\eta}_{\rm res}^{(t)},
		\end{aligned}
		$}
	\label{eq:residual_model}
\end{equation}
where $\boldsymbol{\eta}_{\rm res}^{(t)}$ contains thermal noise and residual PA distortion after cancellation.

\textbf{Step 2: Layered message passing.}
For each WAFT component $k$, the forward mean and variance messages, with marginal variances propagated under a diagonal covariance approximation, are  respectively given by
\begin{equation}
	\scalebox{0.87}{$%
		\begin{aligned}
			\boldsymbol{\mu}_{z,k}^{(t)}
			=
			\mathbf B_k\boldsymbol{\mu}_x^{(t)},
			\,
			\boldsymbol{\nu}_{z,k}^{(t)}
			=
			|\mathbf B_k|^{\circ 2}\boldsymbol{\nu}_x^{(t)},
		\end{aligned}
		$}
	\label{eq:waft_forward}
\end{equation}
where $|\cdot|^{\circ 2}$ denotes element-wise squared magnitude.
After the FTN layer, the corresponding messages are
\begin{equation}
	\scalebox{0.87}{$%
		\begin{aligned}
			\boldsymbol{\mu}_{p,k}^{(t)}
			=
			\mathbf G\boldsymbol{\mu}_{z,k}^{(t)},
			\,
			\boldsymbol{\nu}_{p,k}^{(t)}
			=
			|\mathbf G|^{\circ 2}\boldsymbol{\nu}_{z,k}^{(t)}.
		\end{aligned}
		$}
	\label{eq:ftn_forward}
\end{equation}
For the $k$th WAFT branch, the contributions from the other branches are removed by interference cancellation, i.e.,
\begin{equation}
	\scalebox{0.87}{$%
		\begin{aligned}
			\mathbf r_k^{(t)}
			=
			\mathbf r_{\rm c}^{(t)}
			-
			\sum_{i\in\mathcal K,i\neq k}
			\tilde{\kappa}\mathbf H\boldsymbol{\mu}_{p,i}^{(t)}.
		\end{aligned}
		$}
	\label{eq:branch_ic}
\end{equation}
The equivalent disturbance variance of the $k$th branch is approximated as
\begin{equation}
	\scalebox{0.87}{$%
		\begin{aligned}
			\boldsymbol{\sigma}_{k}^{2,(t)}
			=
			\sigma_{\rm res}^{2,(t)}\mathbf 1_N
			+
			\sum_{i\in\mathcal K,i\neq k}
			|\tilde{\kappa}\mathbf H|^{\circ 2}\boldsymbol{\nu}_{p,i}^{(t)},
		\end{aligned}
		$}
	\label{eq:branch_var}
\end{equation}
where $\mathbf 1_N$ is the all-ones vector, and $\sigma_{\rm res}^{2,(t)}$ denotes the residual disturbance variance after PA cancellation and is estimated from $\hat{\mathbf d}_{\rm eq}^{(t)}$ as 
\begin{equation}
	\scalebox{0.87}{$%
		\begin{aligned}
			\sigma_{\rm res}^{2,(t)}
			&=
			\sigma^2
			+
			\rho_{\rm res}\frac{1}{N}
			\sum_{n=0}^{N-1}\left|[\hat{\mathbf d}_{\rm eq}^{(t)}]_n\right|^2 .
		\end{aligned}
		$}
	\label{eq:res_var}
\end{equation}
Here, $\rho_{\rm res}$ is the residual PA distortion factor.
The resulting branch model is
\begin{equation}
	\scalebox{0.87}{$%
		\begin{aligned}
			\mathbf r_k^{(t)}
			=
			\tilde{\kappa}\mathbf H\mathbf p_k
			+
			\tilde{\mathbf n}_k^{(t)},
			\,
			\mathbf p_k=\mathbf G\mathbf B_k\mathbf x,
		\end{aligned}
		$}
	\label{eq:branch_model}
\end{equation}
where $\tilde{\mathbf n}_k^{(t)}$ denotes the equivalent disturbance of the $k$th branch.
For a discrete linear layer
$\mathbf y=\mathbf A\mathbf u+\mathbf e$, the $m$th entry is
$y_m=\sum_j A_{m,j}u_j+e_m$, where $y_m$, $u_j$, $A_{m,j}$, and $e_m$ denote the corresponding entries of $\mathbf y$, $\mathbf u$, $\mathbf A$, and $\mathbf e$, respectively. Let $u_j\sim\mathcal{CN}(\mu_j,v_j)$ denote the incoming Gaussian message associated with the $j$th variable, and $\sigma_m^2=\mathbb E\{|e_m|^2\}$ denote the disturbance variance. The Gaussian message-passing (MP) projection is defined as $(\boldsymbol{\mu}^{\rm post},\boldsymbol{\nu}^{\rm post})\triangleq\mathcal M(\mathbf y,\mathbf A,\boldsymbol{\mu},\boldsymbol{\nu}, \boldsymbol{\sigma}^2)$, where $\boldsymbol{\mu}^{\rm post}$ and $\boldsymbol{\nu}^{\rm post}$ are the posterior mean and variance of $\mathbf u$, respectively. The update of the $j$th element in $\mathcal M(\cdot)$ is 
\begin{equation}
	\scalebox{0.87}{$%
		\begin{aligned}
			\zeta_{m\to j}
			=
			&\sigma_m^2+
			\sum_{\ell\ne j}|A_{m,\ell}|^2v_\ell,\quad
			v_{j}^{\rm post}=
			\left(
			\sum_m
			\frac{|A_{m,j}|^2}{\zeta_{m\to j}}
			\right)^{-1},\\
			\mu_j^{\rm post}
			&=
			\mu_j
			+
			v_j^{\rm post}
			\sum_m
			\frac{A_{m,j}^*
			\left(y_m-\sum_\ell A_{m,\ell}\mu_\ell\right)}
			{\zeta_{m\to j}}.
		\end{aligned}
		$}
	\label{eq:generic_mp}
\end{equation}
Based on the linear observation model in \eqref{eq:branch_model}, the channel layer updates the posterior mean and variance of $\mathbf p_k$ as
\begin{equation}
	\scalebox{0.87}{$%
		\begin{aligned}
			\left(
			\boldsymbol{\mu}_{p,k}^{{\rm post},(t)},
			\boldsymbol{\nu}_{p,k}^{{\rm post},(t)}
			\right)
			=
			\mathcal M
			\left(
			\mathbf r_k^{(t)},
			\tilde{\kappa}\mathbf H,
			\boldsymbol{\mu}_{p,k}^{(t)},
			\boldsymbol{\nu}_{p,k}^{(t)},
			\boldsymbol{\sigma}_{k}^{2,(t)}
			\right).
		\end{aligned}
		$}
	\label{eq:mh_update}
\end{equation}
With $\mathbf z_k=\mathbf B_k\mathbf x$ denoting the $k$th WAFT branch output, the FTN layer updates its posterior mean and variance as
\begin{equation}
	\scalebox{0.87}{$%
		\begin{aligned}
			\left(
			\boldsymbol{\mu}_{z,k}^{{\rm post},(t)},
			\boldsymbol{\nu}_{z,k}^{{\rm post},(t)}
			\right)
			=
			\mathcal M
			\left(
			\boldsymbol{\mu}_{p,k}^{{\rm post},(t)},
			\mathbf G,
			\boldsymbol{\mu}_{z,k}^{(t)},
			\boldsymbol{\nu}_{z,k}^{(t)},
			\boldsymbol{\nu}_{p,k}^{{\rm post},(t)}
			\right).
		\end{aligned}
		$}
	\label{eq:mg_update}
\end{equation}
Since each WAFT component satisfies the scaled unitary property, i.e.,
\begin{equation}
	\scalebox{0.87}{$%
		\begin{aligned}
			\mathbf B_k^H\mathbf B_k=c_k\mathbf I_N,
			\,
			c_k=\|\mathbf B_k\|_F^2/N.
		\end{aligned}
		$}
	\label{eq:scaled_unitary}
\end{equation}
Let $\mathcal K_{\rm act}=\{k\in\mathcal K:c_k\ge\epsilon_c\}$ denote the nonzero WAFT branch set, where $\epsilon_c$ is a small numerical tolerance. Branches with $c_k<\epsilon_c$ are excluded from the backward update and fusion.
Accordingly, for each $k\in\mathcal K_{\rm act}$, the mean and variance are respectively updated as
\begin{equation}
	\scalebox{0.87}{$%
		\begin{aligned}
			\boldsymbol{\mu}_{x,k}^{(t)}
			=
			{\mathbf B_k^H\boldsymbol{\mu}_{z,k}^{{\rm {post}},(t)}}/{c_k},
			\quad
			\boldsymbol{\nu}_{x,k}^{(t)}
			\approx
			\left|{\mathbf B_k^H}/{c_k}\right|^{\circ 2}
			\boldsymbol{\nu}_{z,k}^{{\rm {post}},(t)}.
		\end{aligned}
		$}
	\label{eq:xk_update}
\end{equation}
The estimates from all nonzero WAFT branches are fused according to their reliabilities. Let $\mu_{x,k,n}^{(t)}$ and $v_{x,k,n}^{(t)}$ denote the $n$th entries of $\boldsymbol{\mu}_{x,k}^{(t)}$ and $\boldsymbol{\nu}_{x,k}^{(t)}$, respectively. The prior variance and mean are
\begin{equation}
	\scalebox{0.87}{$%
		\begin{aligned}
			v_{x,n}^{{\rm pri},(t)}
			=
			\left(
			\sum_{k\in\mathcal K_{\rm act}}
			\frac{1}{v_{x,k,n}^{(t)}}
			\right)^{-1},\,
			\mu_{x,n}^{{\rm pri},(t)}
			=
			v_{x,n}^{{\rm pri},(t)}
			\sum_{k\in\mathcal K_{\rm act}}
			\frac{\mu_{x,k,n}^{(t)}}{v_{x,k,n}^{(t)}}.
		\end{aligned}
		$}
	\label{eq:fuse_var}
\end{equation}
Accordingly, \eqref{eq:fuse_var} performs inverse variance fusion using the marginal branch variances.
Finally, the fused Gaussian message is projected onto the constellation alphabet $\mathcal A$,
leading to the posterior probability
\begin{equation}
	\scalebox{0.8}{$%
		\begin{aligned}
			P^{(t)}(x_n=a)
			=
			\frac{
			\exp\left(
			-\frac{|a-\mu_{x,n}^{{\rm pri},(t)}|^2}
			{v_{x,n}^{{\rm pri},(t)}}
			\right)
			}
			{
			\sum_{b\in\mathcal A}
			\exp\left(
			-\frac{|b-\mu_{x,n}^{{\rm pri},(t)}|^2}
			{v_{x,n}^{{\rm pri},(t)}}
			\right)
			},
			\, a\in\mathcal A.
		\end{aligned}
		$}
	\label{eq:posterior_prob}
\end{equation}
The posterior mean and variance are then updated by
\begin{equation}
	\scalebox{0.87}{$%
		\begin{aligned}
			\hat{\mu}_{x,n}^{(t)}
			=
			\sum_{a\in\mathcal A}aP^{(t)}(x_n=a),
		\end{aligned}
		$}
	\label{eq:post_mean}
\end{equation}
\begin{equation}
	\scalebox{0.87}{$%
		\begin{aligned}
			\hat{v}_{x,n}^{(t)}
			=
			\sum_{a\in\mathcal A}|a|^2P^{(t)}(x_n=a)
			-
			|\hat{\mu}_{x,n}^{(t)}|^2.
		\end{aligned}
		$}
	\label{eq:post_var}
\end{equation}
After collecting $\{\hat{\mu}_{x,n}^{(t)}\}$ and $\{\hat{v}_{x,n}^{(t)}\}$ into $\hat{\boldsymbol{\mu}}_x^{(t)}$ and $\hat{\boldsymbol{\nu}}_x^{(t)}$, respectively, damping is applied as
\begin{equation}
	\scalebox{0.87}{$%
		\begin{aligned}
			\boldsymbol{\mu}_x^{(t+1)}
			=
			\lambda\hat{\boldsymbol{\mu}}_x^{(t)}
			+
			(1-\lambda)\boldsymbol{\mu}_x^{(t)}, \,			\boldsymbol{\nu}_x^{(t+1)}
			=
			\lambda\hat{\boldsymbol{\nu}}_x^{(t)}
			+
			(1-\lambda)\boldsymbol{\nu}_x^{(t)},
		\end{aligned}
		$}
	\label{eq:damp_mean}
\end{equation}
where $0<\lambda\leq 1$ is the damping factor. After $T$ iterations, the final hard decision is obtained as
\begin{equation}
	\scalebox{0.87}{$%
		\begin{aligned}
			\hat{x}_n
			=
			\arg\min_{a\in\mathcal A}
			|\mu_{x,n}^{(T)}-a|^2,
			\, n=0,\ldots,N-1.
		\end{aligned}
		$}
	\label{eq:hard_decision}
\end{equation}
Algorithm~\ref{alg:pa_ic_mlmp} summarizes the overall procedure of the proposed PA-IC-MLMP receiver.
\begin{algorithm}[t]
	\caption{Proposed PA-IC-MLMP Receiver}
	\label{alg:pa_ic_mlmp}
	\footnotesize
	\begin{algorithmic}[1]
		\REQUIRE $\mathbf r$, $\mathbf H$, $\mathbf G$, $\mathbf W(-\alpha,\beta)$, $\{\mathbf B_k\}_{k\in\mathcal K}$, $f_{\rm PA}(\cdot)$, $\gamma_{\rm bo}$, $\tilde{\kappa}$, $\sigma^2$, $\rho_{\rm res}$, $\epsilon_c$, $\mathcal A$, $T$, $\lambda$
		\ENSURE Detected symbol vector $\hat{\mathbf x}$
		
		\STATE Initialize $\boldsymbol{\mu}_x=\mathbf 0$, $\boldsymbol{\nu}_x=\mathbf 1$, $\mathcal K=\{0,1,2,3\}$.
		\STATE Compute the branch powers $c_k=\|\mathbf B_k\|_F^2/N$, $\forall k\in\mathcal K$.
		\STATE Set $\mathcal K_{\rm act}=\{k\in\mathcal K:c_k\ge\epsilon_c\}$.
		
		\FOR{$t=1,\ldots,T$}
		
%		\STATE \textbf{PA layer:}
		\STATE \parbox[t]{0.86\linewidth}{Reconstruct $\boldsymbol{\mu}_s$, estimate $\hat{\mathbf d}$, and obtain $\hat{\mathbf d}_{\rm eq}$.}
		\STATE Cancel PA distortion:
		$
		\mathbf r_{\rm c}=\mathbf r-\hat{\mathbf d}_{\rm eq}.
		$
%		\STATE \textbf{\% Forward message passing: constellation $\rightarrow$ WAFT $\rightarrow$ FTN}
		\FOR{$k\in\mathcal K_{\rm act}$}
		\STATE $\boldsymbol{\mu}_{z,k}=\mathbf B_k\boldsymbol{\mu}_x,\,
		\boldsymbol{\nu}_{z,k}=|\mathbf B_k|^{\circ 2}\boldsymbol{\nu}_x$.
		\STATE $\boldsymbol{\mu}_{p,k}=\mathbf G\boldsymbol{\mu}_{z,k},\,
		\boldsymbol{\nu}_{p,k}=|\mathbf G|^{\circ 2}\boldsymbol{\nu}_{z,k}$.
		\ENDFOR
		
%		\STATE \textbf{Channel layer update and backward message passing}
		\FOR{$k\in\mathcal K_{\rm act}$}
		\STATE $\mathbf r_k=\mathbf r_{\rm c}-\sum_{i\neq k}\tilde{\kappa}\mathbf H\boldsymbol{\mu}_{p,i}$.
		\STATE $\boldsymbol{\sigma}_k^2
		=
		\sigma_{\rm res}^2\mathbf 1_N
		+
		\sum_{i\neq k}
		|\tilde{\kappa}\mathbf H|^{\circ 2}\boldsymbol{\nu}_{p,i}$.
		\STATE \parbox[t]{0.86\linewidth}{$(\boldsymbol{\mu}_{p,k}^{\rm post},\boldsymbol{\nu}_{p,k}^{\rm post})
		=
		\mathcal M(\mathbf r_k,\tilde{\kappa}\mathbf H,\boldsymbol{\mu}_{p,k},
		\boldsymbol{\nu}_{p,k},\boldsymbol{\sigma}_k^2)$.}
		\STATE \parbox[t]{0.86\linewidth}{$(\boldsymbol{\mu}_{z,k}^{\rm post},\boldsymbol{\nu}_{z,k}^{\rm post})
		=
		\mathcal M(\boldsymbol{\mu}_{p,k}^{\rm post},\mathbf G,\boldsymbol{\mu}_{z,k},
		\boldsymbol{\nu}_{z,k},\boldsymbol{\nu}_{p,k}^{\rm post})$.}
		\STATE $\boldsymbol{\mu}_{x,k}
		=
		\mathbf B_k^H\boldsymbol{\mu}_{z,k}^{\rm post}/c_k,\,
		\boldsymbol{\nu}_{x,k}
		=
		|\mathbf B_k^H/c_k|^{\circ 2}\boldsymbol{\nu}_{z,k}^{\rm post}$.
		\ENDFOR
		
%		\STATE \textbf{Constellation layer:}
		\STATE Fuse $\{(\boldsymbol{\mu}_{x,k},\boldsymbol{\nu}_{x,k})\}_{k\in\mathcal K_{\rm act}}$ into $(\boldsymbol{\mu}_x^{\rm pri},\boldsymbol{\nu}_x^{\rm pri})$.
		\STATE Project $(\boldsymbol{\mu}_x^{\rm pri},\boldsymbol{\nu}_x^{\rm pri})$ onto $\mathcal A$ to obtain $(\hat{\boldsymbol{\mu}}_x,\hat{\boldsymbol{\nu}}_x)$.
		
		\STATE \parbox[t]{0.86\linewidth}{Damping update:
		$\boldsymbol{\mu}_x
		\leftarrow
		\lambda\hat{\boldsymbol{\mu}}_x
		+(1-\lambda)\boldsymbol{\mu}_x,$\\
		$\boldsymbol{\nu}_x
		\leftarrow
		\lambda\hat{\boldsymbol{\nu}}_x
		+(1-\lambda)\boldsymbol{\nu}_x.$}
		\ENDFOR
		\STATE $\hat{x}_n=\arg\min\limits_{a\in\mathcal A}|\mu_{x,n}-a|^2,\, n=0,\ldots,N-1$.
		\RETURN $\hat{\mathbf x}$
		
	\end{algorithmic}
\end{algorithm}

\subsection{Computational Complexity}
At each iteration, the complexity of PA reconstruction, distortion propagation, WAFT, FTN, channel, and constellation updates can be approximated as $\mathcal{O}(Q_{\rm os}K_pN)$, $\mathcal{O}(LN)$, $\mathcal{O}(|\mathcal K_{\rm act}|N\log N)$, $\mathcal{O}(\eta_GN)$, $\mathcal{O}(LN)$, and $\mathcal{O}(MN)$, respectively. Here, $Q_{\rm os}$ is the oversampling factor, $K_p$ is the oversampled pulse span used for PA reconstruction, and $\eta_G$ is the effective row sparsity of $\mathbf G$ after FTN correlation truncation, i.e., the number of retained nonzero taps per row. All other taps are set to zero. The channel term is controlled by the physical path number $L$, where $L\ll N$ under the considered sparse channel model. The overall complexity over $T$ iterations is then obtained as
\begin{equation}
	\scalebox{0.87}{$%
		\begin{aligned}
			\mathcal{O}
			\left(
			TN
			\left[
			Q_{\rm os}K_p+2L+|\mathcal K_{\rm act}|\log N+\eta_G+M
			\right]
			\right).
		\end{aligned}
		$}
	\label{eq:30}
\end{equation}
For fixed $T$, the adopted FTN truncation, sparse channel model, and fast Fourier transform (FFT)-type WAFT implementation lead to quasi-linear complexity in $N$.

% ============================================================================
\section{Simulation Results}
\label{sec:sims}
% ============================================================================
This section evaluates the performance of the proposed FTN-WAFT system under nonlinear PA transmission. Unless otherwise specified, the block length is $N=512$, 4-QAM modulation is used, the WAFT chirp parameter is $\beta=7/(2N)$, and the damping factor is $\lambda=0.6$. The AFDM/Nyquist benchmark is given by $(\alpha,\xi)=(1,1)$. The memoryless SSPA model in \eqref{eq:PA} is considered, where $A_{\rm sat}=1$ and $p=2$. In addition, we adopt a three-path channel with equal-power paths, normalized delays in $[0,5]$, and a maximum normalized Doppler shift of $1$. The FTN waveform is generated using a root-raised-cosine (RRC) pulse with roll-off factor $\beta_{\rm RRC}=0.3$, with $Q_{\rm os}=4$, $K_p=8$, $T=15$, and $\rho_{\rm res}=0.25$. All waveforms have the same average power before back-off. The signal to noise ratio (SNR) is defined at the sampled AWGN after receive filtering.

\begin{figure}[t]
	\centering
	\includegraphics[width=0.5\columnwidth]{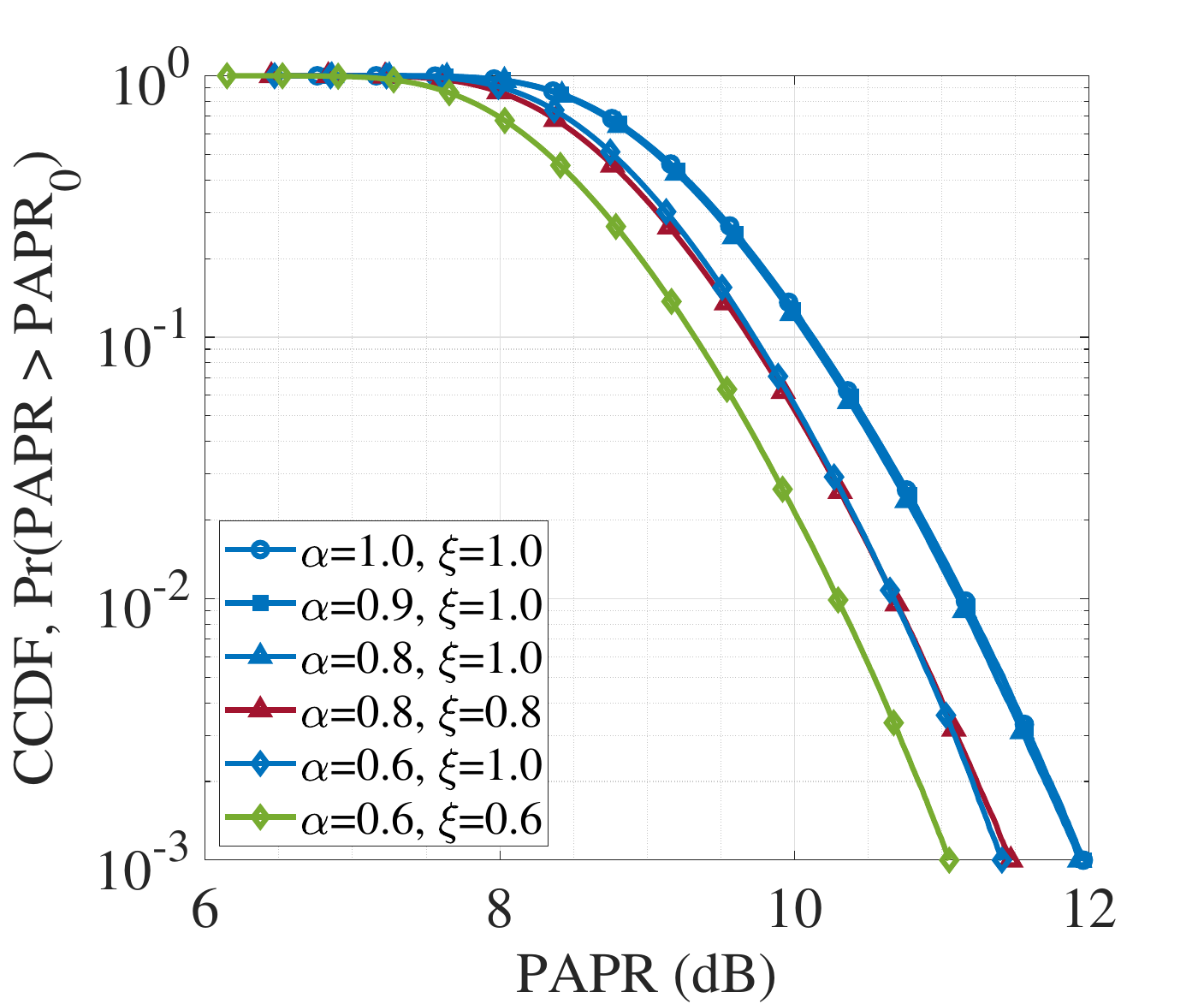}
	\caption{PAPR performance of FTN-WAFT with different parameter settings.}
	\label{fig3}
    \vspace{-0.4cm}
\end{figure}
Fig.~\ref{fig3} compares the complementary cumulative distribution functions (CCDFs) of the PAPR for the AFDM/Nyquist benchmark and the FTN-WAFT waveforms. For the $Q_{\rm os}$-fold oversampled sequence $x_{\rm tx}^{\rm os}[n]$ of $x_{\mathrm{tx}}(t)$, the PAPR is defined as
\begin{equation}
	\scalebox{0.87}{$%
		\begin{aligned}
			{\rm PAPR}
			=
			\frac{\max\limits_{0\le n\le Q_{\rm os}N-1}|x_{\rm tx}^{\rm os}[n]|^2}
			{\mathbb E\{|x_{\rm tx}^{\rm os}[n]|^2\}}
		\end{aligned}
		$}
	\label{eq:papr}
\end{equation}
and the CCDF is ${\rm CCDF}(P_0) = \Pr\{{\rm PAPR}>P_0\}$.
At a CCDF of $10^{-3}$, the PAPR decreases from about $11.96$ dB for $(\alpha,\xi)=(1,1)$ to $11.48$ dB for $(0.8,0.8)$ and $11.04$ dB for $(0.6,0.6)$. 
Thus, joint WAFT/FTN tuning provides additional flexibility for PAPR reduction and adaptive IBO.

\begin{figure}[t]
	\centering		
	\subfigure[Fixed IBO.]{
		\label{FixIBO}
		\includegraphics[width=0.48\columnwidth]{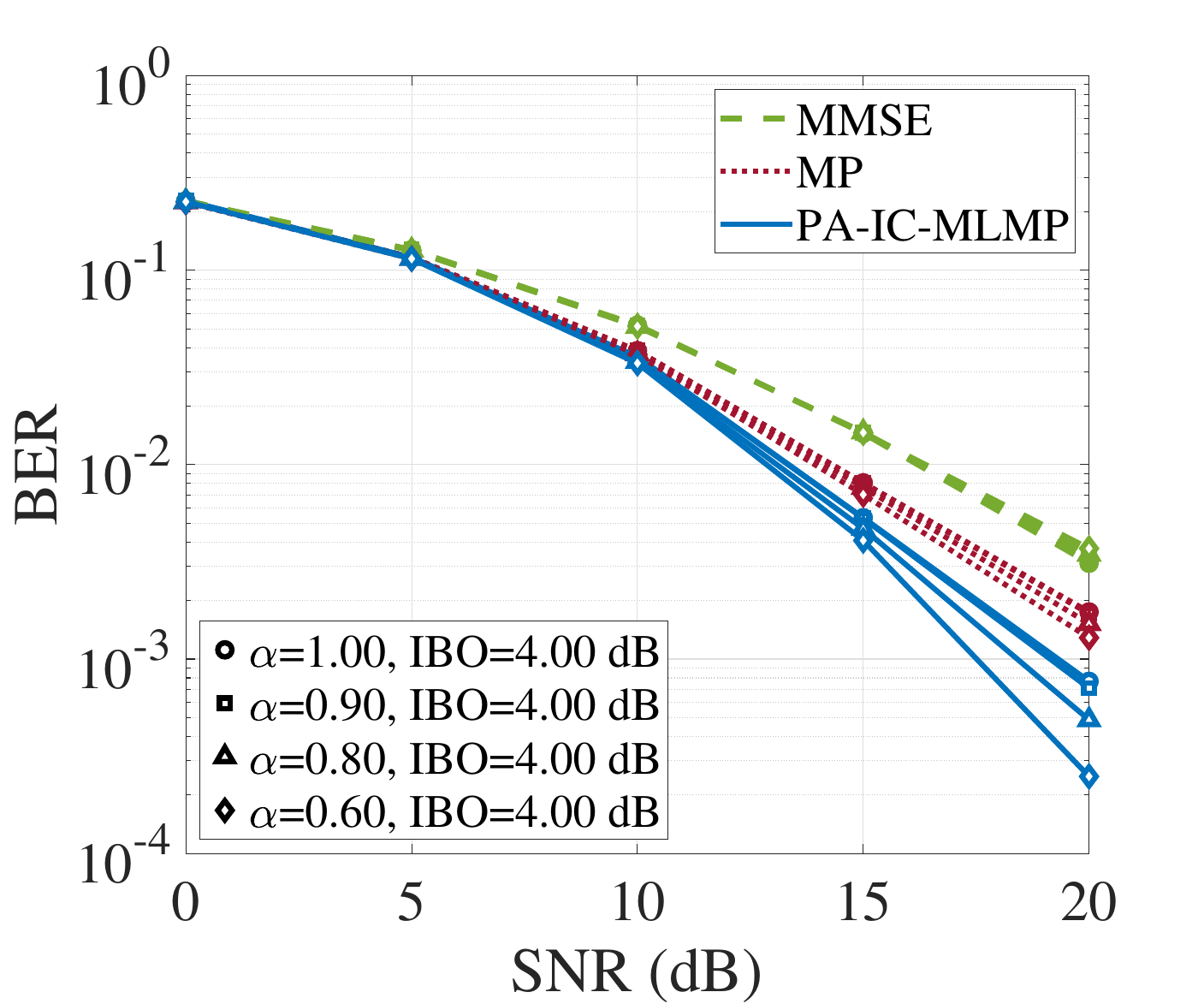}}
	\subfigure[Adaptive IBO.]{
		\label{adaIBO}
		\includegraphics[width=0.48\columnwidth]{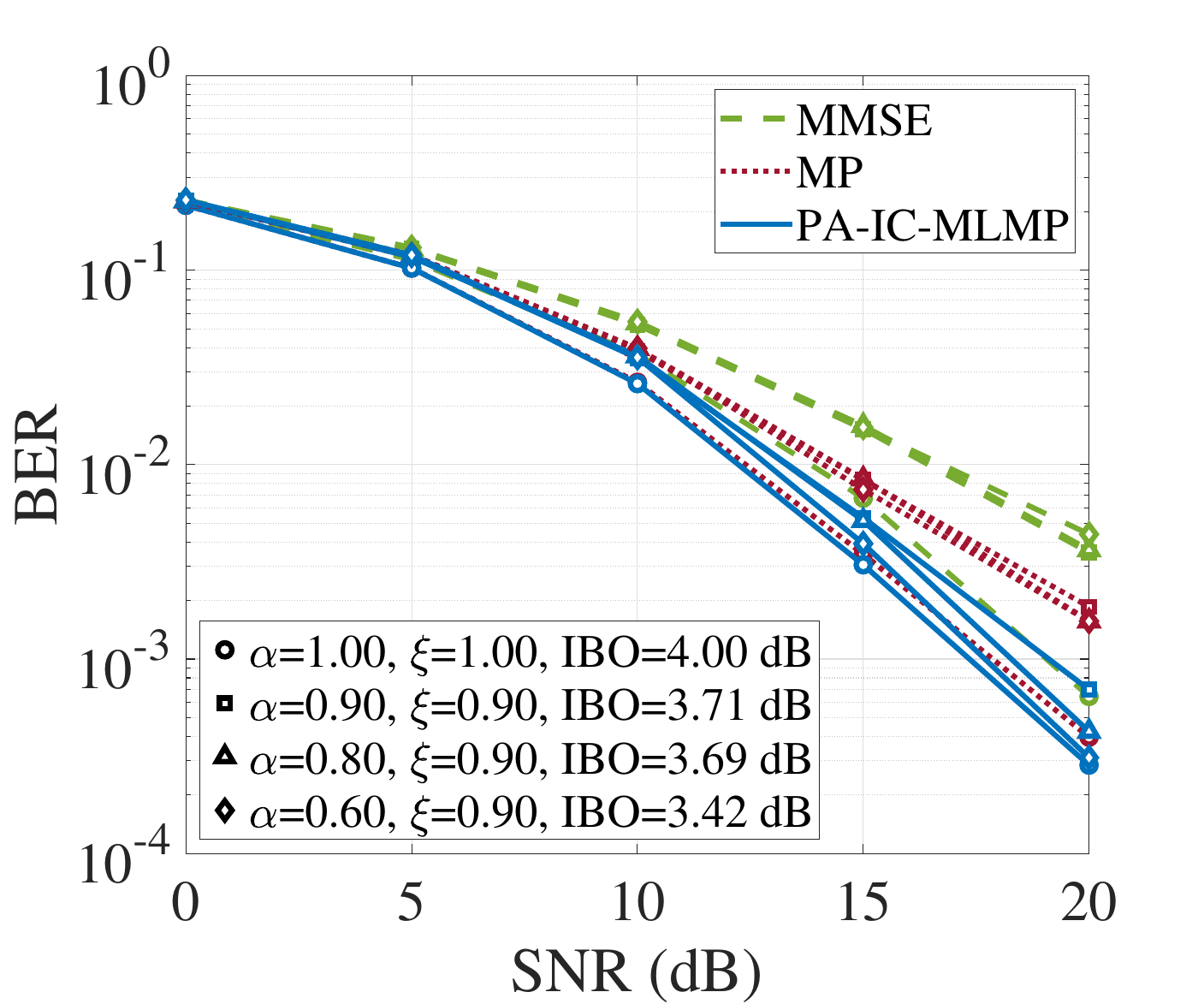}}
	\caption{BER performance versus SNR: (a) fixed IBO WAFT-order comparison and (b) adaptive IBO receiver comparison.}
	\label{fig4}
	\vspace{-0.2cm}
\end{figure}
Fig.~\ref{fig4} compares the BER performance under fixed IBO and adaptive IBO. We set  $\xi=0.9$ and $\mathrm{IBO}=4$ dB  in  Fig.~\ref{fig4}(a) to compare different WAFT orders under the same actual back-off. The MP baseline follows \cite{Raviteja2018OTFSMP} and is adapted for the FTN-WAFT layered structure, while the minimum mean-squared error (MMSE) receiver is used as a linear baseline. The MMSE receiver is degraded by FTN non-orthogonality and PA distortion, while MP improves BER by using the layered signal structure but still treats the residual nonlinear term as an unmodeled disturbance. The proposed PA-IC-MLMP receiver maintains clear BER gains by exploiting PA distortion cancellation and layered message passing. Fig.~\ref{fig4}(b) shows the adaptive-IBO comparison. By jointly adjusting the WAFT order $\alpha$ and FTN compression factor $\xi$, FTN-WAFT exploits PAPR suppression while maintaining robustness over doubly selective channels and controlling FTN-ISI.

\begin{figure}[t]
	\centering
	\includegraphics[width=0.5\columnwidth]{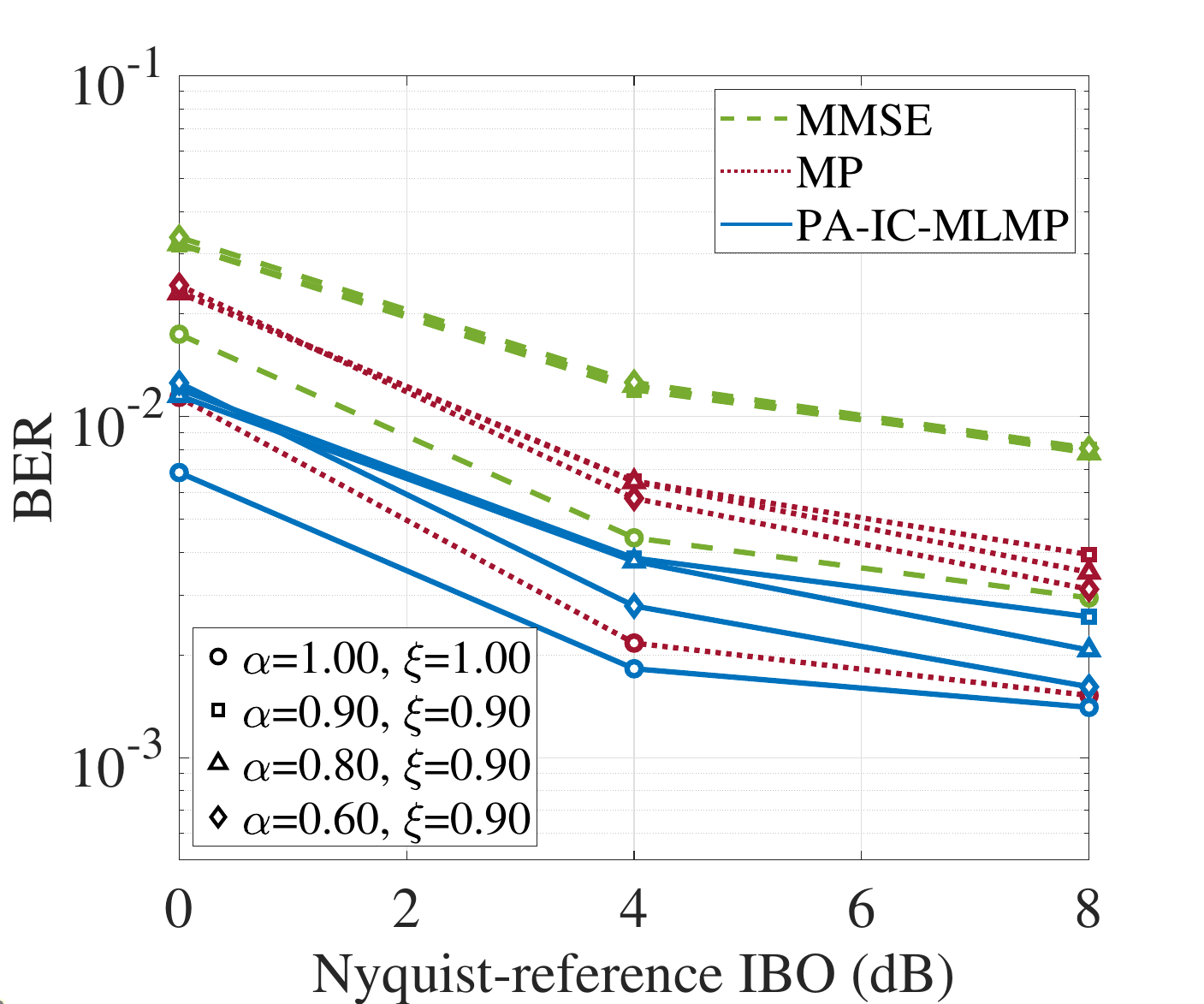}
	\caption{BER performance versus Nyquist-reference IBO at fixed $\mathrm{SNR}=16$ dB.}
	\label{fig5}
    \vspace{-0.4cm}
\end{figure}
Fig.~\ref{fig5} presents BER versus Nyquist-reference IBO at ${\rm SNR}=16$ dB. For the $k$th FTN-WAFT waveform, the actual IBO is ${\rm IBO}_k = {\rm IBO}_{\rm AFDM} - \Delta_{{\rm PAPR},k}$, where ${\rm IBO}_{\rm AFDM}$ and $\Delta_{{\rm PAPR},k}$ are the benchmark IBO and the PAPR reduction at CCDF $10^{-3}$, respectively.
The proposed receiver exhibits improved robustness over most IBO values, especially when nonlinear distortion dominates the performance loss. As the reference IBO increases, the PA approaches its linear operating region and the performance gap among the receivers becomes smaller. This indicates that nonlinearity-aware detection is most beneficial under aggressive IBO reduction.

\begin{figure}[t]
	\centering		
	\subfigure[BER.]{
		\label{ber}
		\includegraphics[width=0.48\columnwidth]{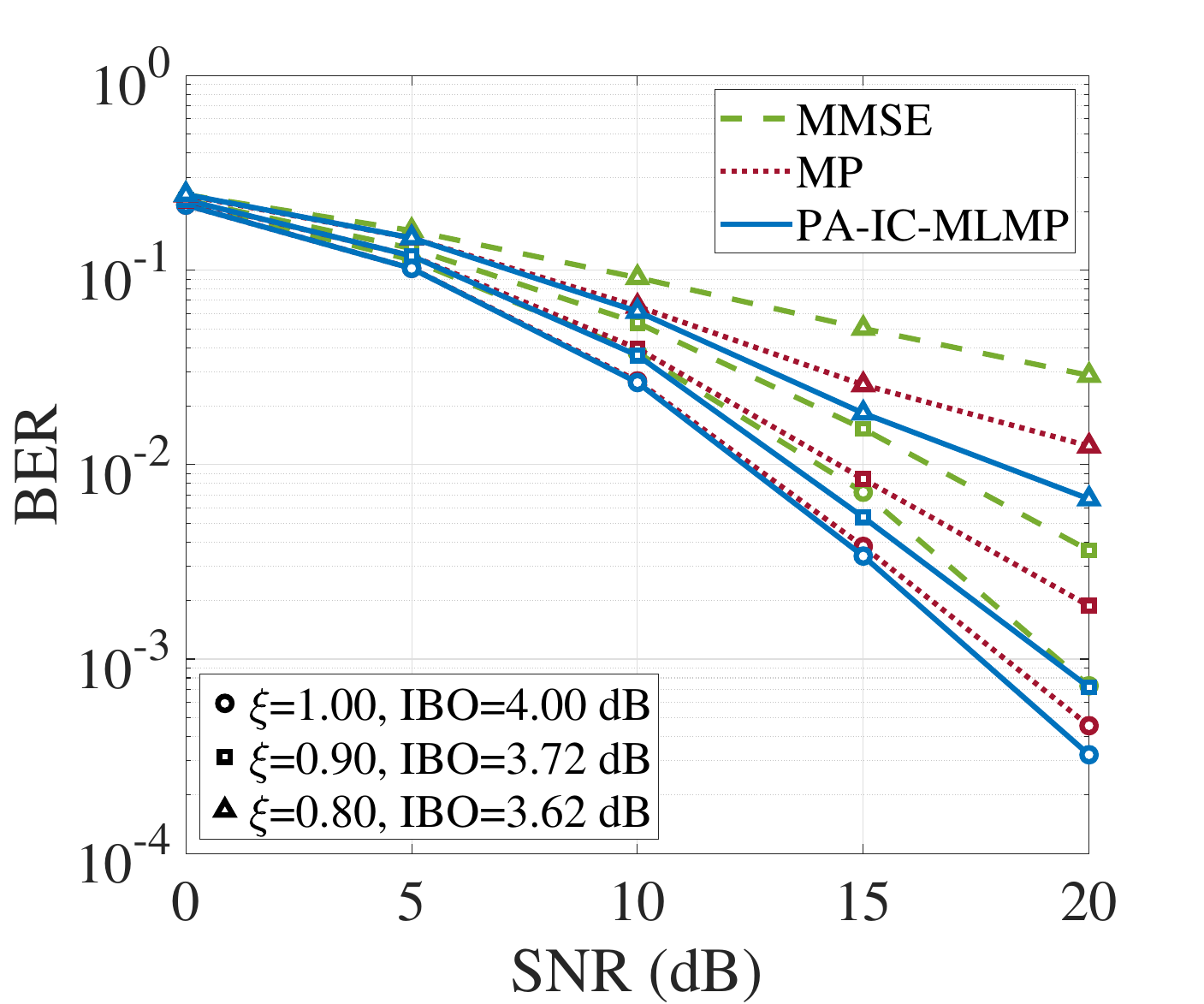}}	
	\subfigure[Effective throughput.]{
		\label{eff_se}
		\includegraphics[width=0.48\columnwidth]{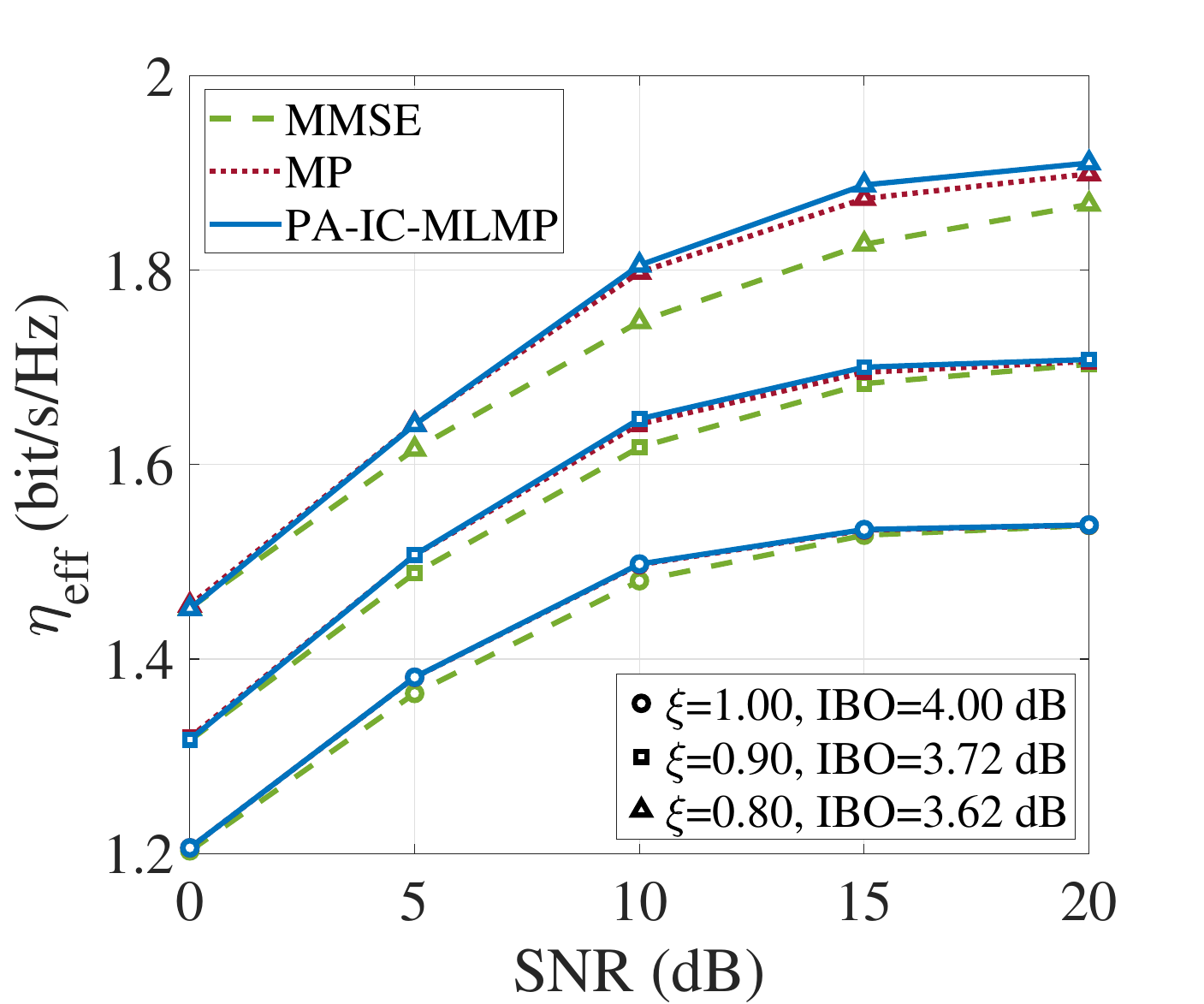}}		
	\caption{BER and effective throughput of FTN-WAFT with different compression factors.}
	\label{fig6}
	\vspace{-0.2cm}
\end{figure}
For the considered uncoded system, we further compare the  effective throughput metric, defined as \cite{yi2025non} 
\begin{equation}
	\scalebox{0.87}{$%
		\begin{aligned}
			\eta_{\rm eff} = \frac{\log_2M} {\xi(1+\beta_{\rm RRC})} (1-{\rm BER}).
		\end{aligned}
		$}
	\label{eq:eta_eff}
\end{equation} 
Fig.~\ref{fig6} shows the BER and effective throughput for $\alpha=0.9$ and
$\xi\in\{1.0,0.9,0.8\}$. It is observed that reducing $\xi$ increases the throughput
gain, but also aggravates FTN-ISI.
The proposed PA-IC-MLMP receiver achieves the lowest BER and the highest effective throughput among the considered receivers.

% ============================================================================
\section{Conclusion}
\label{sec:conclusion}
% ============================================================================

This letter proposed an FTN-WAFT transmission model with nonlinear distortion and a PA-IC-MLMP receiver. By combining PA distortion reconstruction with MLMP, the proposed receiver mitigates PA nonlinearity, FTN-ISI, and channel dispersion with approximately linear complexity scaling. Simulations verified that FTN-WAFT reduces PAPR, supports adaptive IBO, and improves BER and effective throughput under nonlinear PA constraints.

% ============================================================================
%  References
% ============================================================================
\bibliographystyle{IEEEtran}
\bibliography{ftn_waft_references}

\end{document}